\documentclass[11pt,en,cite=super]{elegantpaper}
\pdfoutput=1
\usepackage{algorithm}
\usepackage{algpseudocode}
\usepackage{enumitem}
\newlist{steps}{enumerate}{1}
\setlist[steps, 1]{
label = Step \arabic*:,
 labelsep=8pt,
  labelindent=0.5\parindent,
  itemindent=0pt,
  leftmargin=*,
  before=\setlength{\listparindent}{-\leftmargin},}
\setlist[enumerate]{
  labelsep=8pt,
  labelindent=0.5\parindent,
  itemindent=0pt,
  leftmargin=*,
  before=\setlength{\listparindent}{-\leftmargin},
}
\usepackage{amsmath}
\usepackage{caption}
\usepackage{enumerate}
\usepackage{epsfig}
\usepackage[T1]{fontenc}
\usepackage{float}
\usepackage{graphics}
\usepackage{graphicx}
\graphicspath{{images/}}
\usepackage{subcaption}
\usepackage{csquotes}
\usepackage{array}

\DeclareCaptionFormat{myformat}{#3}
 \title{Continuous Learning and Inference of Individual Probability of SARS-CoV-2 Infection Based on Interaction Data}
 \author{Shangching Liu \\ Synergies Intelligent Systems, Inc.%
 \and Koyun Liu \\  Synergies Intelligent Systems, Inc.%
 \and Hwaihai Chiang \\ Synergies Intelligent Systems, Inc.%
\and Jianwei Zhang$^{\ast}$  \\ 
Universität Hamburg %
 \and Tsungyao Chang$^{\ast}$  \\ Synergies Intelligent Systems, Inc.%
}
\date{Tsungyao Chang, m@sis.ai \\$^{\ast}$Corresponding author: Jianwei Zhang, zhang@informatik.uni-hamburg.de}

\begin{document}

\maketitle

\begin{abstract}
This study presents a new approach to determine the likelihood of asymptomatic carriers of the SARS-CoV-2 virus by using interaction-based continuous learning and inference of individual probability (CLIIP) for contagious ranking. This approach is developed based on an individual directed graph (IDG), using multi-layer bidirectional path tracking and inference searching. The IDG is determined by the appearance timeline and spatial data that can adapt over time. Additionally, the approach takes into consideration the incubation period and several features that can represent real-world circumstances, such as the number of asymptomatic carriers present. After each update of confirmed cases, the model collects the interaction features and infers the individual person's probability of getting infected using the status of the surrounding people. The CLIIP approach is validated using the individualized bidirectional SEIR model to simulate the contagion process. Compared to traditional contact tracing methods, our approach significantly reduces the screening and quarantine required to search for the potential asymptomatic virus carriers by as much as 94\%.
\end{abstract}

\section{Introduction}
The pandemic of the SARS-CoV-2, which causes COVID-19 outbreaks, has a significant impact globally, especially on human life and economic activities. As resources are limited, current policies are having difficulty in identifying and quarantining asymptomatic virus carriers. As a result, it is much harder to control the spread of the virus. To prevent further spread of COVID-19, immediate action is needed. Contact tracing is a method that helps patients recall with whom or where they have been. Identifying contacts and ensuring they do not have a chance to interact with others is critical to slow down the pandemic \autocite{cdc_2019_covid19}.

\begin{figure}[H]
    \centering
    \includegraphics[width=0.5\textwidth]{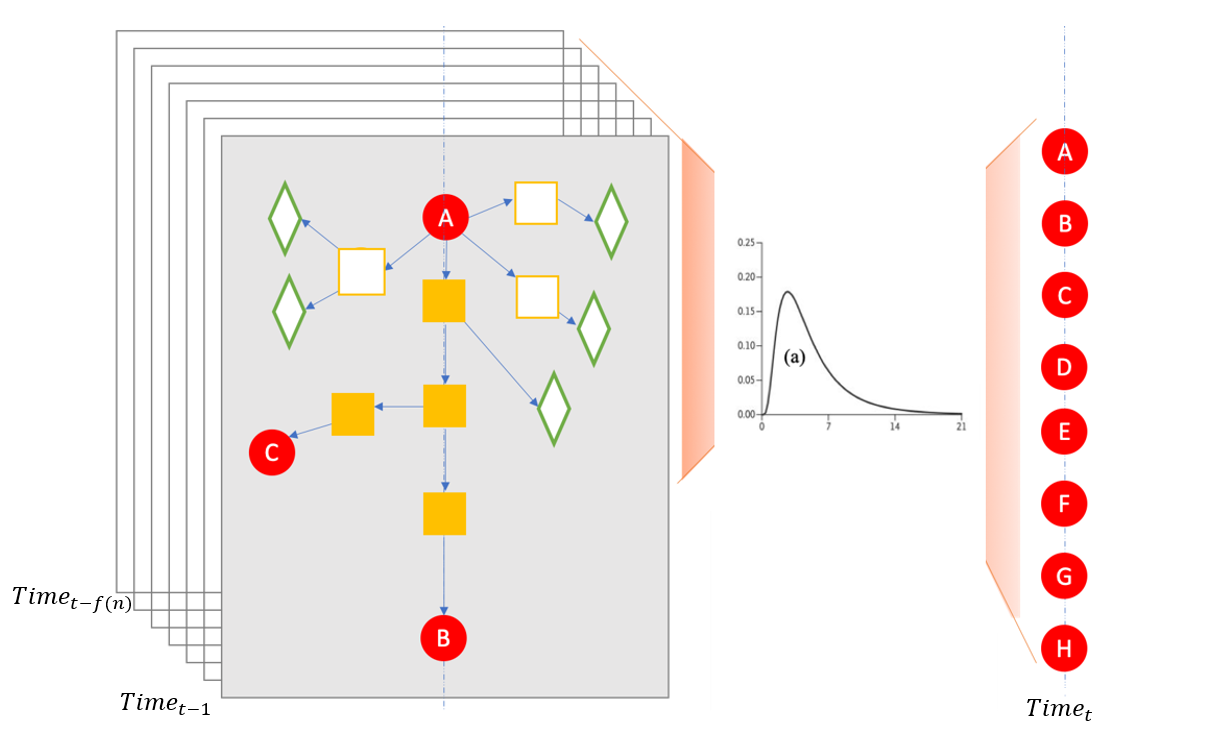}
    \caption{The learning and inference scheme of the  CLIIP model. The diagram shows continuous backtracking learning while the newly infected people are being recorded. Given $n$ is the maximum day of incubation period, the incubation period requires us to group the patients into their actual contagious time according to  the distribution of (a) from time $t$ to range $[t-f(n), t-1]$. The arrangement describes the possible time of the latent infection, which lies in the incubation period of the infected. The inference model of every $t$ is represented by an individual directed graph (IDG) and we use a day as $t$ in our simulation. The red circles denote confirmed infected people, the hollow squares mean exposed people, and a filled square is an individual who stays on the path from one infected person to the others which might be asymptomatic virus carriers. A hollow green diamond is labeled as a healthy person. The arrows denote the possible path of transmission derived from people's location and staying time. The virus will stay in the same place for a while \autocite{van2020aerosol} which makes the last person to leave the specific place run a high risk of getting infected. Also, we define each layer as the number of edges between two nodes. For instance B is a center-surround node in the fourth layer of A.     \label{fig: tracking}}
\end{figure}

This paper is the first in which an approach with continuous learning capabilities is used to analyze the probability of asymptomatic carriers of the severe acute respiratory syndrome coronavirus 2 (SARS-CoV-2). To this end, we compute a ranking model with city GPS spatial dynamics data \autocite{tang2018visual}. The approach is a framework for finding and ranking the source of infection among a moving crowd and can be easily applied to the dynamic modelling of the spreading of the SARS-CoV-2 virus. It is highly efficient in calculating the rich interactive features with continuous data, i.e. it uses continuous time modelling, to approximate the individual probability of being infected since the (Monte Carlo tree search) MCTS on IDG reduces the time to search the important center-surround features. The infection probability of each person exposed in a crowd over time can be quickly obtained by the CLIIP. Moreover, even a superspreader (active in motion, high viral titer, asymptomatic) can be found when we use backward and forward tracking at the same time.  Backward tracking [\ref{fig: tracking}] is the backward finding of a day when someone possibly got infected and forward tracking [\ref{fig: tracking}] means going through the whole day of inference detection on possible days.

\section{Related Work}
Contact tracing is currently the most common way for public health institutions to track infected people and the sources of the virus\ \autocite{contact_eames_2003,contact_Dou_2003}. This method can locate infected individuals and minimize the spread of the virus by isolating them and their contacts at risk  of infection from the public. In past decades, it has been not only used for controlling diseases but also a critical tool for investigating new diseases or unusual outbreaks; for example, SARS and H1N1, two previous pandemics, were suppressed by the help of contact tracing. Governments and health institutes have had or proposed the adoption of contact tracing \autocite{zastrow2020south,contact_istvan_2005,contact_mobile_ajit_2001} to follow the daily routes of residents to decrease the likelihood of infected people's contact with healthy people.

Recently, in order to determine the contact paths of infected people more quickly, the method has been advanced from manual recording and tracking people's mobile phones via Bluetooth \autocite{contact_luca_2020}, or GPS techniques \autocite{cho2020contact, contact_mobile_apple_2020, contact_mobile_Ian_2020, contact_mobile_Justin_2020, contact_mobile_2020}. Moreover, Hellewell et al. \autocite{hellewell2020feasibility} used the model to quantify the potential effectiveness of contact tracing and isolation of the confirmed cases in controlling the outbreak of a severe acute respiratory syndrome coronavirus like SARS-CoV-2. Peng et al. \autocite{Alibaba_health_code_2020} developed the method of a trinary split into red, yellow, or green states to track infected persons. Recent contact tracing methods, such as Zhou et al\ \autocite{zhou2020detecting}, use mobile data with regional infection numbers to predict an individual's possibility to get infected. However, contact tracing cannot identify the probability of asymptomatic carriers and is not always the most efficient method of addressing infectious diseases. Under the current limitation of medical resources, governments can only isolate the people in direct contact with the confirmed cases as the primary way to control the spread of the SARS-CoV-2 virus. 

As the current speed and capacity of virus testing still cannot meet the demand, the outbreak of the COVID-19 is difficult to control. So far, the most feasible way for countries and cities to lessen the spread of infection is to enforce  a lockdown or stay-at-home order to stop unnecessary social interactions of residents. However, the longer lockdown or quarantine has been implemented, the greater its impact on a country's economy, people's mental health and many other aspects of their lives. The non-ranking and exhaustive inspection method of contact tracing with only the confirmed cases is not efficient enough to suppress the outbreak of COVID-19 and its recurrence, especially after the re-opening of a city or country. The detection of asymptomatic infected people, along with appropriate social distancing, effective medical treatments, and the development of vaccination, will greatly determine the extent to which a current or new disease outbreak can be controlled.

As a result, we propose a machine-learning algorithm to predict the spreading of the SARS-CoV-2 virus and reduce the time to locate infected people. We use a gradient boost ensemble learning tree model after the individual state is updated through an IDG to calculate the probability, and continuous learning will keep improving the model of the LightGBM \autocite{ke2017lightgbm} algorithm. It can obtain a better result without parameter adjustment. The CLIIP is an innovative approach combining temporal difference learning which learns by bootstrapping with value function approximation to predict the probability of getting infected when it comes to real circumstances. To continuously measure the real-world physical activity on machine intelligence, the approximation of the value and the professional inference is essential, and our approach bridges the gap between theory and reality.

\section{Methodology}

We develop a framework with the inference model, which is a more efficient and precise method to narrow down the search for potential asymptomatic infected people. It can potentially deduce the source of infection based on the virus infection spreading pathway and the contact tracing process. People's infection paths and their probabilities of infection depend on several critical factors like duration, frequency, and distance of their contact with any infected individual. These factors determine the state of the population infection over time. The continuous learning model based on these phenomena can be used to simulate and analyze the probability of someone's infection status. 

\subsection{Overview}
\begin{figure}[H]
    \centering
    \includegraphics[width=0.7\textwidth]{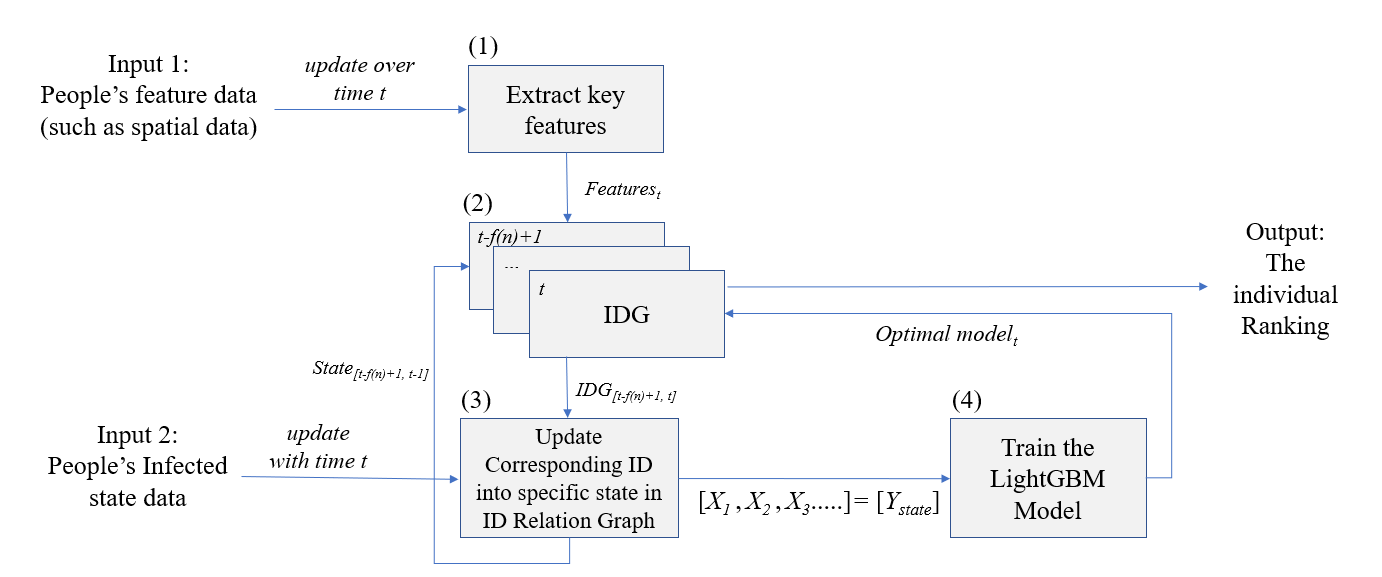}
    \caption{The CLIIP temporal learning framework has two inputs. The first one is continuous spatial data for building an IDG, and the other is a set of labels that provide people's infected states. Combining interaction data with the states and the relation, we can train this model to learn continuously when new data comes into the framework. The updating IDG is built through  comparing the place where two people stop and their overlap time, which defines the relation between two people. An arrow points to the person who stayed longer at a waypoint than the other. According to the path of virus transmission \autocite{dis_henrik_2016}, people's continuous spatial data are a set of essential interaction features, $X_i, i=1,2,3...$, which we use from mobile location data and we can call interaction data.  \label{fig:overview}}
\end{figure}
\begin{itemize}
    \item  Definition of input 1: 
   
    There are $m$ people, ${p=\{h_i(t')\}, i=\{1, 2,...,m\}}$. 
    Then, assume we have k key interaction features of each person at time $t$ to describe people's connection,
    ${h_i(t)=\{h_i^j(t')\}, j=\{1, 2, ..., k\}}$.
    
    In this paper, we use their location and timestamp as their key interaction features. 
    
    \item    
    Definition of input 2: With each time unit, everyone has a label to indicate the state. $\{h_{i}state(t)\}=\{state_1, state_2,..., state_q\}$, where $q$ refers to the number of  people's infected states at time $t$; We use seven kinds of states, which are susceptible \textbf{S}, susceptible\_and\_ quarantined \textbf{Sq}, exposed \textbf{E}, exposed\_and\_ quarantine \textbf{Eq}, infected \textbf{I}, hospitalized \textbf{H}, and recovered \textbf{R}. There is some dependency between these states of a SEIR model \autocite{seir_younsi_2015}.
    

\end{itemize}
The system aims to give out the ranking by order of priority of infection, people between two infected people first then Exposure people as \textbf{E} and then Susceptible people as \textbf{S}, as described in Fig.\  \ref{fig:overview}.
We start from the people's interaction features over time as an input to the framework. The interaction data is filtered out by standard spatial data with more accuracy through map-matching work from Newson et al. work \autocite{newson2009hidden}, or by combining it with other data like credit card transaction data or check-in data as \autocite{check_in_19}. By reconnecting the path for all people, it becomes the social interaction network in the form of an IDG that we use for further research.  
To build up the interaction data as an IDG, we extract the key interaction features describing the dynamic behavior of each person $h_i$ (Fig.\  \ref{fig:overview} step (1)) from
continuous spatial data, from which we can extract the frequency and distance of people's contacts. Another input comes from the SEIR model describing people's state updated each time $t$, like "infected" or "recovered".

\subsection{Combination of SEIR model and interaction data}
\begin{table}[ht]
    \centering
    \begin{tabular}{c|c|c}
      &type       & column      \\ \hline
     1& timestamp & Date Time   \\ \hline
     2& long      & Unique ID   \\ \hline
     3& int       & Device Type \\ \hline
     4& long      & Company     \\ \hline
     5& long      & Device ID   \\ \hline
     6& int       &  Flag       \\ \hline
     7& int       &  Status       \\ \hline
     8& int       &  Event       \\ \hline
     9& location  & Longitude   \\ \hline
     10& location  & Latitude    \\ \hline
     11& location  & encode Longitude   \\ \hline
     12& location  & encode Latitude    \\ \hline
     13& timestamp & Time stamp\\ \hline
    \end{tabular}
    \caption{City Data Set. The resolution of the data is enhanced to one meter in most cases through map-matching \autocite{newson2009hidden}. The data can represent the approximation of human social distance, which is based on Unique ID, Longitude, Latitude and Time stamp.}
    \label{tab:city_data_set}
\end{table}

To prove the effectiveness of the model, we use the dynamic spatial GPS data of a crowd in the city and convert it to approximate the interaction data for 30 consecutive days as input 1 from City GPS spatial data \autocite{tang2018visual}  and Table. \ref{tab:city_data_set}. We calculate the spreading of the virus in the city using the agent-based simulation of the improved SEIR model for SARS-CoV-2 as input 2 to prepare the infection situation.

\begin{figure}[ht]
    \centering
    \includegraphics[width=0.5\textwidth]{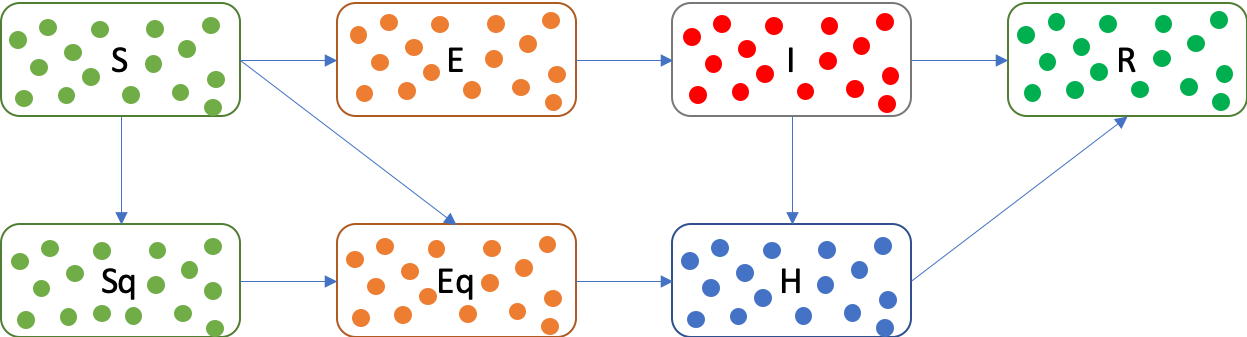}
    \caption{ Extended from the SEIR model for SARS-CoV-2 \autocite{seir_shi_2020}, we propose this model to individualize the contagion process. The original model uses those seven states to separate the people states, and we extend this method to an individual process for further usage.   \label{fig:seir_process}}
\end{figure}

The SEIR model (Definition of Input 2 Fig.\  \ref{fig:overview}) \autocite{seir_younsi_2015} refers to the flows of people between four states: \textbf{S} holds susceptible people, \textbf{E} contains exposed people incubating the disease (and possibly some that are infectious, however, the numbers of infected people are insufficient for the confirmed infected), \textbf{I} holds confirmed infected people, and \textbf{R} recovered people. There are the states, Susceptible quarantined \textbf{Sq}, Exposed quarantined \textbf{Eq}, and Hospitalized \textbf{H}, are taken into consideration as Fig.\  \ref{fig:seir_process}.

With key interaction features from input 1, we generate an IDG at an updated time $t$ (Fig.\  \ref{fig:overview} step (2)), which is a directed acyclic graph used as a people's connection model. We treat each node in the IDG as a person, and each directed edge as a spreading relation between two people who stayed at the same location for a certain time. The direction of the edge means the infection source-destination,  which is defined so that the arrow points to the person who left a place later since he/she is more likely to be infected by the other who left earlier. 
With input 2, we label people's states in the IDG, and update the previous IDG in incubation period $[t-f(n)+1, t-1]$  \autocite{pro_maggie_2018} at the same time (Fig.\  \ref{fig:overview} step (3)). When getting an updated IDG in period $[t-f(n)+1, t-1]$ and $t$, we compute the probability and ranking of each person, including \textbf{S} and \textbf{E}. Using the IDG (Fig.\ \ref{fig:overview} step (4)) and SEIR states we generate each individual's status and calculate the features to feed the model.  The learning process can enhance the capability to search the asymptomatic carriers. Finally, we update the probability and ranking of each person in the period. We then introduce an algorithm using a very simple yet highly efficient searching strategy for training a lightGBM model with data derived from running the SEIR model and relation graph updating. The strength of using the ensemble learning model is that under the real-world scale and computing resources, ensemble learning choosing the computing scale and explainable result. For instance, the searching layers limit on each sampling decision tree. Furthermore, it also allows for easily adding the domain knowledge of public health officials as a feature to deal with uncertainty by their professional insight.

\subsection{Updating states in the IDG}

In the IDG, we label infected people as red nodes, susceptible people \textbf{S} who may be healthy as green nodes, and exposed people who may be infected or virus carriers but not confirmed as yellow nodes. When newly infected people are confirmed from input 2 at time $t$, the incubation period following the distribution in Bays et al \autocite{pro_bays_2020} would apply to each individual. This gives us a way to update states in the IDG between $[t-f(n)+1, t]$, with n being the duration of the incubation period.
  The SEIR model updated every 2 hours between $t$ and $t + 1$, following the step in 4.1 below. Therefore we end up having 530 time-frames of the environment that contain infected persons for 30 days in the city. 
\begin{algorithm}[ht]
  \caption{Algorithm 1: The CLIIP learning algorithm (Fig. \ref{fig:overview})}
  \label{code:recentEnd}
  \begin{algorithmic}[1]
    \While {new\_confirm\_data\_came}
      \If{Initial}
          \State initial\_the\_model(SEIR\_model)
          \State continue
      \EndIf
      \State relation\_graph\_list = []
      \For{Each\_time\_step}
          \State relation\_graph = update(IDG)
          \State relation\_graph\_list.append(relation\_graph)
          \State SEIR\_model = update\_SEIR\_model(relation\_graph\_list)
          \State new\_model = update\_lightGBM\_when\_SEIR\_updated(SEIR\_model)
          \If{new\_model}
                \State process\_list = []
                \If{Graph.find\_infected\_in\_first\_layer}
                    \State put\_each\_person\_on\_path\_into\_process\_list()
                    
                    \Comment{The inference to collect candidates of asymptomatic carriers.}
                \EndIf
                \State add\_E\_state\_to\_process\_list()
                \State add\_S\_state\_to\_process\_list()
                \State give\_out\_ranking\_by\_lightGBM(process\_list) 
                \Comment give out probability of being infected
          \EndIf
      \EndFor
    \EndWhile
  \end{algorithmic}
\end{algorithm}

After that, we use a continuous learning algorithm from Algorithm\ \ref{code:recentEnd} \autocite{CLIIP2020Code} to build the CLIIP model and the LightGBM model by using a set of IDG before time $t$ as a training data set and the SEIR state as a label. In each time-step, the relation graph is updated in the algorithm by updated IDG to form or change the relationship between nodes. Simultaneously, the SEIR is updated by the next time point. Then the CLIIP approach starts calculating the important individual surrounding features such the contact time of infected people. We assume the $node_{path}$ is the person on the path between two infected people, $node_{\textbf{E}}$ is the exposure person and $node_{\textbf{S}}$ is the susceptible person. The measurement of importance is by the order of $node_{path}$, $node_{\textbf{E}}$, and $node_{\textbf{S}}$. The process of counting all surrounding features is simple regarding the collection of the training dataset, but it costs too much. Based on the nodes in the graph having their weights and the probability of asymptomatic carriers, we speed up by performing the search based on the Monte Carlo tree search (MCTS) \autocite{mcts_edward_1959,mcts_dijkstra_1959} method to get the surrounding information of the nodes with no-repeat ID searching.

\subsection{Ranking process}
\begin{figure}[ht]
    \centering
    \includegraphics[width=0.5\textwidth]{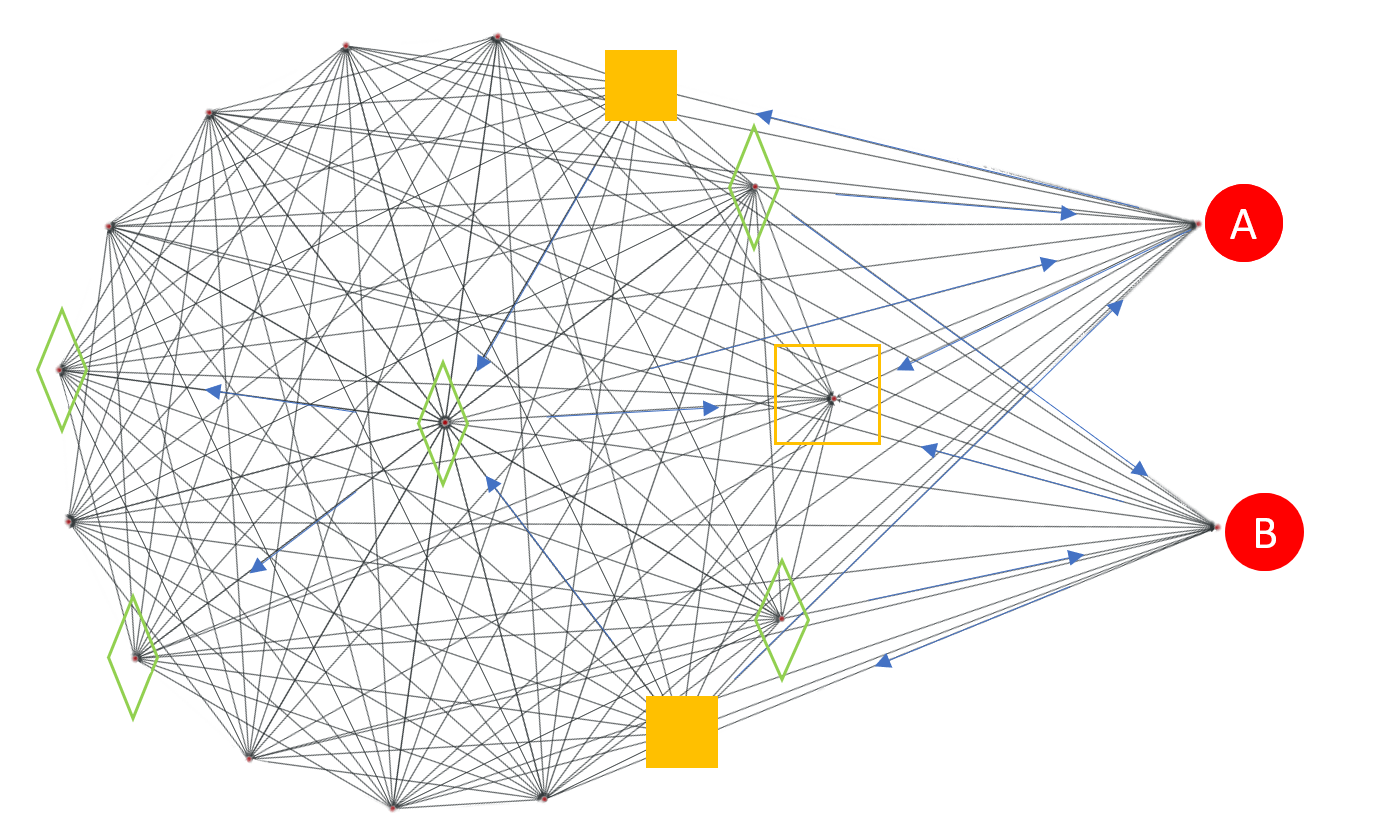}
    \caption{One CLIIP result of the small cluster within one day of IDG as an example. The forward tracking is based on the connection of people, and each edge is directed toward other nodes. We highlight the important arrows to explain this tracking method. The bigger squares depict a higher probability of getting infected, like the one in the middle of the cluster. \label{fig:graph_tree}}
\end{figure}

If we know the new people who got infected from input 2, we backtrack the route of transmission by using incubation period to begin searching in the range of $[t-f(n), t-1]$ days ago. Then we do the forward tracking as shown in Fig.\  \ref{fig:graph_tree}. If we find the source of the virus in the first layer, the search will stop, and we will rebuild all relations of IDG. Then we will start to predict the possibility of people in the order that \textbf{E} goes first and then \textbf{S}. Else if we see it in other layers of the search, we put the people between the path of the first group and add people of status \textbf{E}, which is not in the infected route, into the second group, and collect \textbf{S} to the third group. The ordinal numbers of groups  are the ranking order for calculating the probability by the LightGBM model. The input interaction features of the model will be [$X_{\Delta Time}$, $X_{\Delta Distance}$, $X_{Infected people\_around}$, $X_{Exposed\_around}$], the annotation between (3) and (4) in Fig. \ref{fig:overview}. $X_{\Delta Time}$ and $X_{\Delta Distance}$ is the duration and closest distance between two IDs inside the data. The other interaction features $X_{Infected people\_around}$ and $X_{Exposed\_around}$ stand for several infected people, and exposed people around them. The label Y is the state generated from the SEIR model. The use of this interaction feature is motivated by the inference logic that a virus must come from the people around an infected person. And the output is labeled from the SEIR model simulation. The following Fig. \ref{fig:graph_tree} is a computed example for the result.

Finally, for each person at a specific time, we can both have their infected states and calculate the probability of asymptomatic carriers.

\section{Implementation of individualized SEIR model}

\subsection{SEIR model updating steps}

    

The epidemic data used in this paper comes from \autocite{seir_shi_2020}. Due to the limited data we assume that there are 100 infected people in the group. Then the other states are the same ratio as in \autocite{seir_shi_2020}, except for \textbf{S}; that is the number of ID recorded in the data set being assigned to other states. We initialize parameters: $S=13331$; $E=889$;  $I=100$; $Sq=358$; $Eq=64$; $H=164$; $R=4$. As \textbf{E} people we use the possible list being in contact with the initial infected people \textbf{I}. To extend the distribution of the SEIR model into the individual scale, we follow  the steps below. The update process should be based on the real interaction data as IDG.
\begin{steps}
    \item Load new model in next time-step
    \item If (member of  $S_t$ > member of $S_{t-1}$) get $\Delta S$ from $Sq_{t-1}$ to $S_t$
    \item If (member of $Sq_t$ > member of $Sq_{t-1}$) get $\Delta Sq$ from $S_{t-1}$ to $Sq_t$
    \item If (member of $E_t$ > member of $E_{t-1}$) get $\Delta E$ from $possible\_list\_of\_E$ generated by $I_{t-1}$ with relation built by certain 
    $\Delta Time$ and $\Delta distance$.
    \item If (member of $Eq_t$ > member of $Eq_{t-1}$) get $\Delta Eq$ from $\{possible\_list\_of\_E -E\}$
    \item If (member of $I_t$ > member of $I_{t-1}$) get $\Delta I$ from $E_{t-1}$ and the probability of choice depends on individual incubation period of $E_{t-1}$. 
    \item If (member of $H_t$ > member of $H_{t-1}$) get $\Delta H$ from $\Delta Eq_{t-1}$ and $I_{t-1}$ by random.
    \item If (member of $R_t$ > member of $R_{t-1}$) get $\Delta R$ from $I_{t-1}$ and $H_{t-1}$ by random. (This should improve by depending on a curved day)
\end{steps}
The IDG from Section 3 is the foundation to build the $possible\_list\_of\_E$ inside of the update state. 

\subsection{Simulation and model building}
To simulate the situation more realistically, we made some arrangements regarding the initial individuals. First, we randomly sorted people into the \textbf{Eq}, \textbf{Sq} group. For state \textbf{I}, we split 100 initial values into two groups; one group was chosen randomly, the other group was selected depending on the first and second layers of the first group of people. This process could yield the primary connection between the first group of infected people. Then the state \textbf{E} people will be picked from a group of connection to state \textbf{I}. Then the rest of the people will become \textbf{S}. Then we applied the update rule to the last section. From here, we could get people's state as $X$ input, for which now we merely considered the interactive time, interactive distance, first, second, and third layers of infected people, and exposed numbers. Moreover, we attributed label $Y$ in a specific state to the IDG. This will be updated by future data. After each update of the model, we use 3:7 as the ratio of the test data set and the training data set. Then we get the relation between the history record and interactive contact.
\section{Results}
\begin{figure}
\begin{minipage}{.5\textwidth}  
\centering
\includegraphics[width=\textwidth]{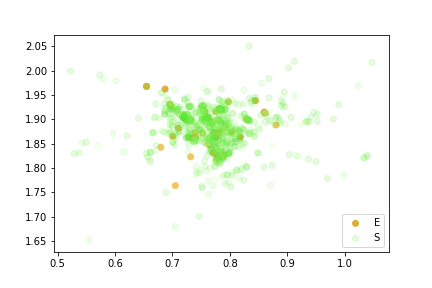}
\captionof{figure}{City ranking overview. The x\_axis is the 
longitude, and the y\_axis is the latitude of the position.
Then we illustrate the probability of all people of an \textbf{S} (Susceptible) and \textbf{E} (Exposed) state of ranking in the 8\% proportion of the center of the city area by the intensity of the colors based on the last record of location from interaction data in Fig.\ \ref{fig:graph_tree}. }
\label{fig:ranking}
\end{minipage}\hfill
\begin{minipage}{.45\textwidth}
    \centering 
\begin{tabular}{lc} \hline
     & Average AUC of our model   \\ \hline
     score & 0.96 \\ \hline \hline
     & Feature importance\\ \hline
    & [$X_{\Delta Time}$, $X_{\Delta Distance}$, $X_{I}$, $X_{E}$] \\ 
    result& [0.48, 0.13, 0.22, 0.23 ]\\
      \hline
    \end{tabular}%
\captionof{table}{The average AUC score. This is accurate enough to find the principle of our SEIR model, the ranking model of all data, on the condition that we eliminate the people with state \textbf{I} that do not have any surrounding features measured in up to three layers. The prediction divided people into two groups by probability of $0.5$: either they will or will not be infected. The result of feature importance shows the contact time to the high-risk location which is more relevant for judging whether someone has been infected or not, and where $X_I$ and $X_E$ are the total numbers of state \textbf{I} and \textbf{E} in the first three surrounding layers, respectively.}
\label{tab:acc}
\end{minipage}
\end{figure}

With this infected situation model, we create a perfect fit in the individual SEIR model. First, we used the incubation period time distribution to begin searching in the range of the previous 5-7 days. Then we continued the finding process until infected people were found and started ranking the people on the path. In the real world, things become more complex as the virus could spread out not only from people's contacts. This will entail more missing nodes on the disease spread map which will require us to consider more layers on this condition. However, we can claim that the method can locate about 96\%\  (Table. \ref{tab:acc} average AUC of our model)\ asymptomatic people in the group of people if we have all their surrounding label records and transfer data. The results of crowd ranking visualization shows the ranking distributed from susceptible people to infected people and indicates the probability (darker means high probability) of asymptomatic virus carriers in Fig.\ \ref{fig:ranking}.  



 We estimate the average precision of the model and by seeing the interaction feature importance in Tab.\   \ref{tab:acc}, we claim that the model found the rules inside of the SEIR model such as the transmission rule. 

\begin{figure}[H]
    \centering
     \includegraphics[width=0.6\textwidth]{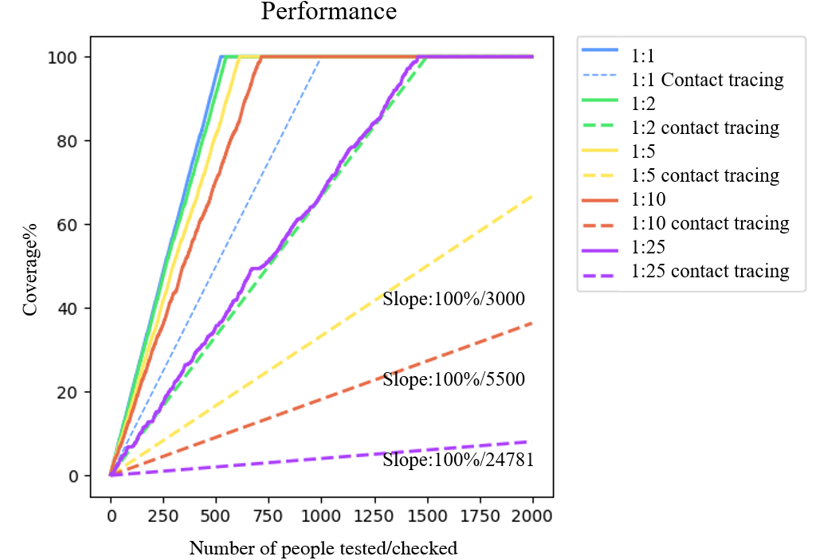} \caption{Comparison with average contact tracing on different scales between infected people and non-infected people. The base unit of the ratio is 500. The ratio shows the comparisons between infected people and non-infected people. The dashed line refers to the performance of ordinary contact tracing, and the solid line of the same color means the result obtained by the proposed method. If the testing follows the order of probability generated by our approach for a bigger group of people than the purple lines demonstrate, $\frac{1500}{24781} \approx 6\% $ of people need to be tested in comparison with contact tracing, which means reducing up to 94\% of screening resources usage. }
\label{fig:comparison}
\end{figure}

As resources are limited in the real world, there should be some priority in ranking the crowd. As in Fig.\  \ref{fig: tracking}, the first level of people exposed to infected people has higher priority than the second level and so on. 
Fig.\  \ref{fig:comparison} shows the correlation between the CLIIP model and the baseline of contact tracing. We use a different group of samples to demonstrate the results. The base unit of a ratio is 500 people, so the blue line shows the 1000 people group, and there are 500 already recognized as infected people. In contrast, the blue dash line shows the primary contact tracing performance that compares to the method. Generally, it needs to search until the end to make sure no person out there has been missed.

Moreover, we use this model to test a larger group of people with more healthy people in the test group. The CLIIP model can cover most infected people when checking the same number of people because of our ranking order, which speeds up testing. We plot more than thousands of points to address the result. Furthermore, the baseline we compare is the average performance of contact tracing. Thus the CLIIP model can find infected people more precisely and decrease the required social and medical resources.

\section{Conclusions}

We propose a novel interaction-based inference learning approach whose major advantage lies in calculating the individual probability of getting infected from interactions along a timeline. In each phase of the pandemic the government could use people's interactive data to generate the risk to individuals, which would enable the government to stop the spreading of the virus.  

In addition, the learning algorithm allows us to employ multi-modal datasets and interactive features such as weather, subjective feelings of individuals, wearing masks \autocite{mask_dekai_2020}, hand washing, and other health-related factors. Using the mask wearing probability distribution, we will be able to find the approximated situation of the real world by putting relevant factors into our approach under enough real world interactive data. This could further increase the accuracy in calculating and ranking the infection probability. Our approach can be further applied to more real world scenarios:
\begin{itemize}

\item \textbf{Precisely identifying and predicting the most likely virus carriers}
    
    Ranking the probability of potential asymptomatic carriers of the crowd by our approach helps with precisely controlling the spread of the SARS-CoV-2 virus. This approach simulates very well under the condition of sufficient spatial mobile data during citywide outbreaks. Healthcare officials can develop a more precise control or quarantine strategy toward the affected regions, areas, or individuals than a citywide lockdown. Furthermore, an adaptive and flexible "exit" strategy can also facilitate the re-opening and maintain normal economic activities with a limited quarantine. 
   
    \item \textbf{Searching for superspreaders}
      The disease spreading map in our IDG makes the ranking of superspreaders possible. Following the state of contacting people, the superspreaders are most likely to be in the path between two infected people, which is key to suspending the spread of the virus. Using our approach to analyze individuals of the surrounding layer of the spreader, the possibility of being a superspreader can be described as the equation below:
\begin{equation}
possibility = layer_1\_infected * w_1 + layer_2\_infected * w_2 *...layer_n\_infected * w_n,
\end{equation}
    which guides the search for superspreaders and creates more learning samples for further finding action. This enhances the learning precision and accelerates the inference process significantly.
    \item \textbf{Decision support for saving resources}
    
    The approach can simulate the situation after executing the policy. The individual model of the virus spread could give suggestions on: 
    \begin{itemize}[--- ]
    \item \textit{Disinfection, sterilization, and preservation.} 
    
    Based on the distribution of the spread probability in outdoor areas, indoor simulation is also possible. Combining surveillance camera data with the CLIIP to give the infected index of each contacting region, the precise disinfection of, for instance, elevator buttons in the area is possible, for example,  when a threshold of the accumulative possibility of being touched by high-risk people is reached.
     
    \item \textit{Optimal testing times.}  
     
   With new testing methods like nucleic acid tests, PCR based tests, antigen tests, and serology tests, we could add the features of fail testing probability and recalculating the individual infection probability to our approach. Considering all individuals in society based on our approach, it is possible to calculate the R0, a mathematical term that indicates how contagious an infectious disease is. However, we need to rebuild the model of the CLIIP and make labels like R0 to train the new model. Decision makers can refer 
  to the R0 to obtain the infection degree of an area and thus decide on the testing times and methods.

    \end{itemize}

    \item \textbf{Exogenous reinfection}
    
    To counter the reinfection of SARS-CoV-2, the CLIIP can reuse the data from the first infection model to predict the probability of reinfection for the rest of the people. Although the source of exogenous people is unclear in the transmission route, especially regarding the latency of SARS-Cov-2, our approach is able to observe each person in society to calculate the individual probability of reinfection.

\end{itemize}
\bigskip
 Beyond the virus spreading,  our approach can be applied to modelling, learning and inference on the individual level of general latent influence networks, such as in P2P e-commerce, searching for terrorists, predicting risks of digital security and so on. 
 In social networks, for example, people send  diverse comments to each other, influencing the others via their mood, intent, and thus generating the individualized relation graph.
By measuring people's center-surrounded commented mood/intent and by continuous learning, their decision-making policy on issues such as purchasing behaviours, finding terrorism and preventing digital virus spreading, and so on, can be gradually modelled and their future actions can be precisely predicted.




\nocite{
seir_shi_2020, 
zastrow2020south,
contact_mobile_2020,
contact_Dou_2003, 
contact_luca_2020, 
contact_mobile_Ian_2020, 
contact_mobile_Justin_2020, 
contact_eames_2003,
contact_istvan_2005, 
contact_mobile_ajit_2001, 
contact_mobile_apple_2020, 
pro_bays_2020, 
seir_younsi_2015,
pro_maggie_2018,
mcts_edward_1959,
mcts_dijkstra_1959,
dis_henrik_2016,
Alibaba_health_code_2020,
ke2017lightgbm,
tang2018visual,
hellewell2020feasibility,
cho2020contact,
ke2017lightgbm,
tang2018visual,
zhou2020detecting,
newson2009hidden,
van2020aerosol,
cdc_2019_covid19,
CLIIP2020Code}
\newpage
\printbibliography

\appendix

\typeout{get arXiv to do 4 passes: Label(s) may have changed. Rerun}

\end{document}